\documentclass[letterpaper,11pt]{article}
\usepackage{graphicx}
\usepackage[margin=1in]{geometry}
\usepackage{paper}
\hypersetup{pdftitle={The Determinants of Net Interest Margin in the Turkish Banking Sector: Does Bank Ownership Matter?}}

\newcommand{\bib}{bibliography.bib}

\title{
  {\fontsize{23pt}{26pt}\selectfont % Custom size between \huge and \Huge
  The Determinants of Net Interest Margin in the Turkish Banking Sector: Does Bank Ownership Matter?}%\thanks{The views expressed are those of the authors and do not necessarily reflect the views of any institution.}%
}
\author{
  \sc Fatih Kansoy%
  \thanks{Contact: \href{mailto:fatih.kansoy@economics.ox.ac.uk}{fatih.kansoy@economics.ox.ac.uk} or \href{mailto:kansoy@gmail.com}{kansoy@gmail.com} } \\
}
\date{2012}
\begin{document}
\maketitle

% Call the available command to redefine the plain page footer.
\available{https://www.fatih.ai/nim.pdf}
%\available[]{https://www.fatih.ai/cbdc.pdf}
\thispagestyle{plain}

{\setstretch{1}
\begin{abstract}

\noindent    This research presented an empirical investigation of the determinants of the net interest margin in Turkish Banking sector with a particular emphasis on the bank ownership structure. This study employed a unique bank-level dataset covering Turkey`s commercial banking sector for the 2001-2012. Our main results are as follows. Operation diversity, credit risk and operating costs are important determinants of margin in Turkey. More efficient banks exhibit lower margin and also price stability contributes to lower margin. The effect of principal determinants such as credit risk, bank size, market concentration and inflation vary across foreign-owned, state-controlled and private banks. At the same time, the impacts of implicit interest payment, operation diversity and operating cost are homogeneous across all banks.

\medskip


\medskip


\medskip

\noindent \textbf{To cite:} Kansoy, Fatih. (2012). The Determinants of Net Interest Margin in the Turkish Banking Sector: Does Bank Ownership Matter? \textit{Journal of BRSA Banking and Financial Markets}, \textbf{6}(2), 13-49.
\end{abstract}
}

% % BODY OF THE DOCUMENT
% \clearpage
% \pagebreak 
 \newpage

% \tableofcontents

\onehalfspacing

\section{Introduction} \label{sec:intro}
Many studies have found that financial development and efficiency have strong relations with economic conditions. \cite{levine1998stock} showed that the financial dynamics play a positive role on economic growth. Also, \cite{calderon2003direction} concluded that improvement in the financial system brings economic growth.  This is especially true for Turkey where banking sector has been improving for ten years after the 2001 banking crisis with parallel its  GDP growth rate. For instance, as far as our calculation, from the last quarter of 2001 to the first quarter of 2012, the ratio of total banking market assets to GDP increased by 51 percent. It means that banks started to play a dominant and increasingly significant role in the financial system of Turkey. 

In the recent credit crunch faced by the world, global financial giants announced substantial losses and some of them went bankrupt or were nationalised. In contrast to this, the Turkish banking sector declared considerable profits \citep{aysan2010macroeconomic}.  This situation is interpreted as a success of the Turkish Banking sector and also triggered a fundamental discussion on the efficiency of the Turkish banking industry \citep{ft2010}. In this case, the Net Interest Margin\footnote{Net Interest Margin is defined as net interest income over total assets}  (hereafter NIM) as a significant component of the efficiency and the profitability of the banking sector needs to be investigated and understood with its major determinants in Turkey, in order to gain a clear perspective \citep{demirgucc1999determinants}. 

Although other studies have dealt with the competition structure, performance and profitability of the Turkish banking sector (for example, see \cite{aysan2010macroeconomic}) or some Turkish banks who have been included in several cross-country studies (see \cite{demirgucc1999determinants}; \cite{aysan2010macroeconomic}; \cite{kasman2010consolidation}) which examined the NIM on the Turkish banking system using the static estimation model. The relatively little research or no special research on the determination of NIM in Turkey has created a gap in the NIM literature. Therefore, the principal motivation of this study is to fill this gap. This research, to the best of knowledge, will provide the first dynamic estimation of the NIM and take the structure of bank ownership into account in particularly. Furthermore, the previous studies used limited samples of the Turkish banking sector which are available on international databases such as the BankScope. However, this study uses rich and specific dataset for every single bank and also the data sample of this study is based on seasonal data covering the whole Turkish banking sector while previous studies mostly used yearly datasets. 

% \texttt{indentfirst} package.

\subsection{\textbf{Research Questions}}
\subsubsection{What are the key determinants of the Net Interest Margin in the Turkey Banking sector?}

The contemporary literature suggests that the determinants of NIM are numerous and differ across the regions, countries and even the structure of ownership. The determinants of NIM can be divided into three parts I) bank-specific, II) industry-specific, and III) macroeconomic specific.  Some studies claim that the macroeconomic determinants have the most crucial effects on the determination of the NIM, whereas a substantial number of studies argue that the bank-specific and industry-specific factors are very important factors that affect the margin. Therefore, to answer or investigate this question is vital to gain a clear perspective for the NIM in Turkey. It is important because understanding the drivers of the NIM is valuable both from a macro and micro view \citep{liebeg2006determinants}. From a macro or an economic stability perspective, it is helpful for a monetary authority to understand whether the increasing or decreasing NIM is mainly attributable to the microeconomic factors or the macroeconomic conditions. For example, if one of the crucial components of the NIM is determined by the volatility of nominal interest rate instead of the competition structure of the banking sector, the government authority ought to focus on how to provide a stable macroeconomic environment to decrease the cost of financial intermediation services. On the contrary, if the main element of NIM is market power, the public policy should be aimed at promoting competition in the banking sector. Regarding the micro vision, specifying the main foundations behind the moving (widening or tightening) of the NIM might enable investors to evaluate potential changes in the NIM in Turkey. For these reasons, this study targets to offer a better understanding of the elements that drive the NIM and to contribute some noteworthy policy insights regarding the macro and micro perspectives.

\subsubsection{Are intercepts and the effects and coefficients of the determinants the same for all ownership structures?}

In the literature, the source of interest revenue and the costs vary by bank ownership structure. The ownership structure of bank may play different a role on performance or profits of bank. Consequently, the strategies and incentives related to the NIM might differ by the bank ownership. For instance, \cite{drakos2003assessing}, \cite{mody2004foreign}, and \cite{williams2007factors} claimed that the foreign ownership has a negative significant effect on the NIM, whereas,\cite{demirgucc1999determinants}, \cite{liebeg2006determinants}, \cite{fungavcova2011determinants} have found a positive relationship. On the contrary, \cite{claessens2001does} and  \cite{dabla2007bank} argued that there is no significant relationship between the ownership structure and the NIM. 

On the other hand, such discussions assume that the coefficient of determinants and their effects on the interest margin are the same for all different bank structures. However, in this research, it is assumed that the intercepts and the coefficient of the determinants may differ for all banks with different ownership form.  Therefore, the aim of answering above question is to make a contribution to this controversial debate by investigating the Turkish Banking sector. 
 	
The rest of the study will be organised as follows; Section II reviews the existing literature. Section III describes data sources and discusses the variables and provides the empirical model and methodology. Section IV provides empirical results. Section V consists of the robustness test and its results and finally the last section concludes.

\section{Literature Review}

The contemporary literature of the determinants of the NIM has been elaborated in \cite{ho1981determinants} pioneering study. In their dealership model, they assume that a bank is a risk averse dealer in the loan market and it acts as an intermediary between fund lenders and the borrowers. Their theoretical model claims that the NIM is mainly contingent upon four main factors: the risk aversion degree, the market structure, the banking transaction magnitude and the divergence of the interest rate on credits and deposits.

Ho and Saunders (1981) set a two-step estimation approach using 100 main US banks from 1976 to 1979 for seasonal periods. In the first step of estimation, authors regress the individual banks` NIM against the banks specific characteristics such as risk aversion, and the implicit interest payment. The second step involves the estimation of pure spread that is explained by the market structure and macroeconomic variables. \cite{lerner1981} criticised the model of \cite{ho1981determinants}.  He argued that the model fails to consider the potential heterogeneity across the banks. \cite{maudos2004factors} responded by extending the dealership model. They incorporated the operating cost into the initial model and supplied an elaborative explanation of the relation between the riskiness and the NIM. In particular the new model distinguishes between sector risk and loan risk in addition to separating potential factors that affect the NIM. 

The empirical studies in the literature attempting to analyse the bank interest margin vary on a large number of countries sample (see \cite{demirgucc1999determinants}) to a single country examples (see Fungacova and Poghosyan, 2011). Also some studies examine particularly developed countries \citep{maudos2009determinants} and the emerging countries \citep{mody2004foreign} moreover; there are also some regional studies like the Central Eastern European countries \citep{claeys2008determinants}.

\subsection{\textbf{Studies on Developed Countries}}

Studies focusing on developed countries for the determinations of NIM are generally parallel with the theoretical structure of the \cite{ho1981determinants} model.\cite{angbazo1997commercial} by using the US data for the period 1989-1993 added the credit risk and the interest risk into the model. This study indicates that the interest margin has a negative relation to the liquidity and competition, whereas positive relation in the case of management quality, market power and gross income volatility. Similarly, \cite{saunders2000determinants} apply the two step model for the US banking system and the bank data for the six European countries for the 1998-95 period. Results show regulatory issues and macroeconomic conditions have crucial effects on the NIM for the banking sectors of those countries.

\cite{maudos2004factors} make a very influential contribution to the NIM literature. As mentioned above, \cite{maudos2004factors}  have responded to the critics of \cite{lerner1981} on the pioneering \cite{ho1981determinants}'s model by expanding the theoretical model through including operating cost as a determinant of the interest margin with their empirical study for the five European countries banking industries in the period 1992-2000. \cite{maudos2004factors} claim that the banking intermediation is reflected by the operating cost as a function of the deposit taken and credit granted. For this reason they conclude that the banks have to cover their operating cost by charging higher interest margin. Except the operating cost, they also conclude that interest rate and credit risk, capital adequacy, implicit interest payment and management efficiency have a positive relationship with the NIM.

Similar to \cite{maudos2004factors} and \cite{valverde2007determinants} have made a significant contribution to the original model. They improve the model by incorporating both conventional and non-conventional operations of the bank in order to observe the impact of diversification on the NIM by considering multi-output model for seven European countries. The evidence of their dynamic estimation model suggests that the specialisation in non-conventional operations induces a narrowing in margin and a widening in the market share as a result of cross-subsidisation. Finally, their results show a negative relation between the GDP growth and interest margin.

\cite{hawtrey2008bank} carry out another study on fourteen OECD countries, the According to the study of Hawtrey and Liang`s (2008) the bank interest margin is negatively affected by management quality and positively affected by the credit risk and implicit interest margin. Whereas, \cite{williams2007factors} claims that there is a negative relation between credit risks and the NIMs in Austrian banking industry. Some selected studies on developed countries for the NIM is provided in Table 8.

\subsection{\textbf{Studies on Developing Countries, Regions and a Large Sample of Countries}}

The empirical studies on the determination of the NIM in developing countries have controversial results compared to developed countries. For this reason Brock and Rojas-Suarez (2000) argued that the generally true methods for developed countries cannot be valid for less developed countries. For example, covering a large sample of countries around the world \cite{demirgucc1999determinants} have analysed the determinants of NIMs by employing bank level data for 80 countries for the years 1988-95 and they have found out that the effect of the banking ownership on the NIM is different for developed countries compared to the developing ones. They claim that in developed countries, domestic banks realise higher interest margins than foreign banks, in contrast in the developing countries foreign banks realise higher margins than the domestic banks. Their evidence suggests that macroeconomic and regulatory factors have substantial effects on the interest margin. 

In parallel with \cite{demirgucc1999determinants} results, the study of \cite{mody2004foreign} for seven Latin American countries show that the foreign banks in these countries are able to exhibit lower margins and also lower the cost down to less than the cost of domestic banks. This is also suggested by \cite{drakos2003assessing} for the Central Eastern European (CEE) Countries. In contrast to \cite{drakos2003assessing}; \cite{liebeg2006determinants} argue that foreign banks apply higher spread than the domestic banks, even though their work is on the CEE Countries. \cite{claeys2008determinants} examine the effects of macroeconomic environment, industry specific features and bank specific characteristics on the NIM in the CEE countries for the years 1994-2001.

\cite{maudos2009determinants} improve the model with their study on the Mexican banking industry and their results suggest that the operating costs and liquidity ratio have a positive and significant effect on the NIM. Another country-based study is on Russia, carried out by \cite{fungavcova2011determinants} and they provide the first evidence on the determinants of NIM in terms of the bank ownership effect. Their findings show that the level of margin varies over domestic, public and foreign banks.  \cite{gounder2012determinants} show that NIM is positively associated with the implicit interest margin, operating cost and the market share, whereas the management quality and liquidity ratio is negatively related. Table 9 and Table 10 provide some information for the studies on the interest margin of developing countries and international cross-countries. 

\section{Empirical Approach and Data}

\subsection{\textbf{Data Information}}

This study uses three different types of data for analysing the NIM of the Turkish Banking System. The first one is the bank-specific data, which was obtained from the Banks Association of Turkey\footnote{The Banks Association of Turkey is a professional organisation, which is a legal entity with the status of a public institution, established pursuant to Article 19 of the Banks Act. The purpose of the Association is to represent the rights and interests of the banking sector and to work for the growth and healthy functioning of the banking system, and strengthening of competition power and preventing unfair competition in the market and to develop the banking profession in Turkey. (Obtained from http://www.tbb.org.tr/ )} . The second one is that the industry-specific data which reflects the main features of the Turkish Banking industry. This study uses some market-specific data, such as Herfindahl Index, and it was obtained from the Banks Association of Turkey. The last data illustrates the macroeconomic environment of Turkey at a particular period, such as the real GDP growth and inflation. The Macroeconomic data was obtained from the Central Bank of Republic of Turkey. To eliminate the direct impact of the 2000 and 2001 economic and banking crises, the quarterly dataset of this study ranges from the last quarter of 2001 to the first quarter of 2012 and includes twenty-three commercial banks.

This dataset has three major advantages when compared to the most previous studies. First, the dataset covers almost all of the commercial banks in the sector, in contrast many previous studies have used the Bankscope dataset, which has a selection bias since the Bankscope dataset includes only the main players and excludes the small players. Secondly, the dataset of this study consists of quarterly data, not annually, and this allows us to interpret changes over four quarters. Lastly, all banks in the dataset use the same accounting and regulatory regime and the same type of balance sheet. These advantages prevent potential distorting influence in the analyses. Table-\ref{tab1sum} and \ref{tab2corr} give the data statistical summaries and the cross correlation matrix information, respectively.

% Table generated by Excel2LaTeX from sheet 'Sheet9'
\begin{table}[H]
  \centering
  \caption{~ Cross Correlation Matrix}
  \resizebox{\textwidth}{!}{    \begin{tabular}{l|rrrrrrrrrrrrr}
          & NIM   & RA    & RBD   & OC    & LOGTA & LQR   & \multicolumn{1}{l}{MNGMT} & IIP   & DPZTG & \multicolumn{1}{l}{DVRSTY} & HHI   & GDP   & INF \\
    \midrule
    \midrule
    NIM   & 1     &       &       &       &       &       &       &       &       &       &       &       &  \\
    RA    & 0.302 & 1     &       &       &       &       &       &       &       &       &       &       &  \\
    RBD   & 0.053 & -0.068 & 1     &       &       &       &       &       &       &       &       &       &  \\
    OC    & 0.307 & 0.483 & 0.054 & 1     &       &       &       &       &       &       &       &       &  \\
    LOGTA & -0.127 & -0.474 & 0.023 & -0.362 & 1     &       &       &       &       &       &       &       &  \\
    LQR   & -0.033 & 0.187 & -0.159 & 0.060 & -0.188 & 1     &       &       &       &       &       &       &  \\
    MNGMT & -0.073 & -0.069 & 0.075 & 0.263 & -0.085 & 0.018 & 1     &       &       &       &       &       &  \\
    IIP   & 0.660 & 0.148 & 0.077 & 0.300 & -0.116 & -0.108 & 0.072 & 1     &       &       &       &       &  \\
    DPZTG & -0.095 & 0.128 & -0.026 & 0.106 & -0.071 & 0.092 & 0.011 & -0.077 & 1     &       &       &       &  \\
    DVRSTY & 0.009 & 0.032 & 0.066 & 0.124 & -0.089 & 0.124 & -0.146 & 0.064 & 0.077 & 1     &       &       &  \\
    HHI   & 0.116 & 0.117 & 0.387 & 0.236 & -0.338 & 0.034 & 0.001 & 0.039 & 0.020 & 0.092 & 1     &       &  \\
    GDP   & -0.140 & -0.035 & -0.087 & -0.007 & -0.016 & -0.007 & 0.035 & -0.067 & 0.019 & 0.016 & 0.118 & 1     &  \\
    INF   & 0.227 & 0.082 & 0.470 & 0.198 & -0.263 & 0.078 & -0.041 & 0.104 & -0.010 & 0.081 & 0.661 & -0.246 & 1 \\
    \bottomrule
    \bottomrule
    \end{tabular}}%
  \label{tab2corr}%
\end{table}%

\begin{table}
  \centering
  \caption{~Summary Statistics }
    \begin{tabular}{l|lccccc}
    \multicolumn{1}{l}{\textbf{Variable}} &      & \textbf{Mean} & \textbf{Std. Dev.} & \textbf{Min} & \textbf{Max} & \textbf{Observations} \\
\hline
    NIM   & Overall & 1.488 & 3.496 & -15.630 & 73.190 & N =    966 \\
          & Between &       & 0.604 & 0.519 & 3.617 & n =      23 \\
          & Within &       & 3.446 & -14.661 & 74.159 & T =      42 \\
    RA    & Overall & 14.028 & 7.972 & -3.270 & 91.610 & N =     966 \\
          & Between &       & 4.917 & 8.998 & 25.655 & n =      23 \\
          & Within &       & 6.356 & -3.727 & 82.970 & T =      42 \\
    RBD   & Overall & 7.485 & 13.532 & 0.000 & 125.370 & N =     966 \\
          & Between &       & 6.147 & 0.000 & 27.476 & n =      23 \\
          & Within &       & 12.122 & -17.081 & 116.299 & T =      42 \\
    OC    & Overall & 1.359 & 1.818 & -7.440 & 27.390 & N =     966 \\
          & Between &       & 0.769 & 0.614 & 4.371 & n =      23 \\
          & Within &       & 1.655 & -10.452 & 24.378 & T =      42 \\
    LOGTA & Overall & 6.776 & 0.807 & 4.390 & 8.230 & N =     966 \\
          & Between &       & 0.755 & 5.679 & 7.876 & n =      23 \\
          & Within &       & 0.326 & 5.487 & 7.567 & T =      42 \\
    LQR   & Overall & 35.133 & 19.966 & 2.900 & 271.930 & N =     966 \\
          & Between &       & 14.599 & 17.443 & 80.808 & n =      23 \\
          & Within &       & 13.948 & -23.481 & 283.895 & T =      42 \\
    MNGMT & Overall & 60.844 & 97.955 & -1498.740 & 1831.510 & N =     966 \\
          & Between &       & 19.099 & 33.810 & 118.447 & n =      23 \\
          & Within &       & 96.156 & -1556.343 & 1794.468 & T =      42 \\
    IIP   & Overall & 0.471 & 4.196 & -52.580 & 82.450 & N =     966 \\
          & Between &       & 1.588 & -4.450 & 5.587 & n =      23 \\
          & Within &       & 3.898 & -47.659 & 77.334 & T =      42 \\
    DPZTG & Overall & 26.868 & 477.392 & -99.900 & 14266.000 & N =     966 \\
          & Between &       & 81.311 & 3.003 & 387.596 & n =      23 \\
          & Within &       & 470.715 & -451.038 & 13905.270 & T =      42 \\
    DVRSTY & Overall & 0.381 & 0. 873648 & -6.471 & 12.210 & N =     966 \\
          & Between &       & 0.250 & 0.293 & 0.898 & n =      23 \\
          & Within &       & 0.838 & -10.298 & 7.205 & T =      42 \\
    HHI   & Overall & 10.882 & 0.643 & 9.940 & 12.230 & N =     966 \\
          & Between &       & 0.000 & 10.882 & 10.882 & n =      23 \\
          & Within &       & 0.643 & 9.940 & 12.230 & T =      42 \\
    GDP   & Overall & 5.092 & 5.865 & -14.740 & 12.590 & N =     966 \\
          & Between &       & 0.000 & 5.092 & 5.092 & n =      23 \\
          & Within &       & 5.865 & -14.740 & 12.590 & T =      42 \\
    INF   & Overall & 14.845 & 14.977 & 4.350 & 70.370 & N =     966 \\
          & Between &       & 1.446 & 8.214 & 15.147 & n =      23 \\
          & Within &       & 14.910 & 4.049 & 70.069 & T =      42 \\
    \hline
     \end{tabular}%
  \label{tab1sum}%
\end{table}%

\subsection{\textbf{Empirical Strategy}}
\subsubsection{Static Panel Estimations}

At first, this study starts with Pooled Ordinary Least Squares (POLS) estimation for static model. The OLS estimators are assumed that they are consistent when all independent variables are not correlated with the error term. However, the fact that this assumption can be violated in the case of that there are unobserved bank specific impacts or independent variables might be correlated with the error term for example, endogeneity problem.
The empirical model of this study for conventional cross-section regression is as follows:

\begin{equation} \label{eq1}
    NIM_{i,t}= \xi + \delta'\textbf{X}_{i,t} + \mu_{i} + \varepsilon_{i,t}
\end{equation}

where  $NIM_{i,t}$ it is the NIM of bank $i$ at time $t$, $\xi$ is a constant term, $\mu_{i}$  is an independently distributed error term with $E[\varepsilon_{i,t}]=0$ also $ \mu_{i}$ is an unobserved bank specific effects which is not correlated the error term.  $\textbf{X}$ represents the set of independent variables as follow
\begin{equation}\label{eq2}
    \textbf{X}_{i,t} = \mathlarger{\sum_{k=1} ^{K}} \beta_{k} PIM_{k,i,t} + \mathlarger{\sum_{l=1} ^{L}} \Psi_{l} BS_{l,i,t} + \mathlarger{\sum_{m=1}^{M}}\lambda_{m}MME_{m,t}
\end{equation}
$\beta_{k}$ is the $K$ coefficients of the pure interest rate margin (PIM), $\Psi_{l}$  is the $L$ coefficients of the bank specific (BS) determinants and $\lambda_{m}$  is the $M$ coefficients of the market specific and macroeconomic specific determinants that are constant over all banks in a given time.
 
When performing the POLS regression, this study does not take into unobserved bank specific effects account for the model-\ref{eq1}. Hence, heterogeneity of the bank specific might be appearing of the estimated parameters. For these reasons, this study estimates the model incorporates unobserved bank specific effects by Fixed and Random Effect methods. Combining the bank specific effects has many advantages. For instance, it permits accounting for specific effects. After that, in order to decide between POLS and Random Effect as an estimation method the Breusch and Pagan`s LM test is used. 

\begin{itemize}
    \item $H_{0}$: Irrelevance of unobserved bank specific effects.
    \item $H_{A}$: Relevance of unobserved bank specific effects.
\end{itemize}

Rejecting the null hypothesis implies that the POLS is not proper method for estimation and vice versa. Also, to test the misspecification between the Random Effect and Fixed Effect methods the Hausman test is used. All these tests can be seen at Table-\ref{tabbplm} and \ref{tabhaus}

\subsubsection{Dynamic Panel Estimation}

The static models are not able to investigate the potential dynamism. To capture the tendency of the NIM and to be persistent over time this study considers that the current values of the NIM might be determined by their previous values \citep{valverde2007determinants}. This study therefore estimates the following dynamic model, with the lagged dependent variable among the regressors. In dynamic framework, this study`s model can be re-written in the following form

\begin{equation}
\label{eq3}
    NIM_{i,t}= \xi + \Psi_{1}NIM_{i,t-1} + \delta'\textbf{X}_{i,t} + \mu_{i} + \varepsilon_{i,t}
\end{equation}

for $i=1, \dots, N$ and $t=2, \dots, T$ where it has the standard error component structure;

\begin{equation}
\label{eq4}
\begin{aligned}
  E[\mu_{i}] = 0 \\
    E[\varepsilon_{i,t}] = 0 \\
      E[\varepsilon_{i,t}\mu_{i}] = 0 
\end{aligned}
\end{equation}

for $i= 1, \dots, N$ and $t=2, \dots,T$

In order to eliminate bank specific effect the first difference is taken;

\begin{equation} \label{eq5}
    NIM_{i,t} - NIM_{i,t-1} = \xi + \Psi_{1}\Big(NIM_{i,t-1} - NIM_{i,t-2} \Big) + \delta'\Big(\textbf{X}_{i,t}-\textbf{X}_{i,t-1}\Big)  + \Big(\varepsilon_{i,t} - \varepsilon_{i,t-1}\Big)
\end{equation}

The lagged dependent variable $NIM_{i,t-1} - NIM_{i,t-2}$  and the error term $\varepsilon_{i,t} - \varepsilon_{i,t-1}$ are correlated with each other which indicates that the explanatory variables are likely endogenous. The econometric presumptions indicate that the error term is not serially correlated and the explanatory variables are weakly exogenous. Thus, the moment conditions based upon difference estimator is employed by dynamic GMM estimator for Equation-\ref{eq3}

\begin{equation} \label{eq6}
  \mbox{\Large E}\bigg( NIM_{i,t-k}\Big(\varepsilon_{i,t} - \varepsilon_{i,t-1}\Big) \bigg) = 0 \quad  for \quad  t=3, ... T, k \geq 2
\end{equation}

\begin{equation} 
\label{eq7}
    \mbox{\Large E}\bigg(\textbf{X}_{i,t-k}\Big(\varepsilon_{i,t} - \varepsilon_{i,t-1}\Big) \bigg) = 0 \quad  for \quad  t=3, ... T, k \geq 2
\end{equation}

This can be written the matrix presentation as;

\[
  \textbf{K}=
  \left[ {\begin{array}{ccccccc}
   y_{i,1} & 0 & 0 & \cdots & 0 & \cdots & 0\\
   0 & y_{i,1} & y_{i,2} & \cdots & 0 & \cdots & 0\\
   \vdots & \vdots & \vdots & \vdots& \vdots & \vdots & \vdots \\
   0 & 0 & 0 & \cdots & y_{i,1} & \cdots & y_{i,T-2}\\
  \end{array} } \right]
\]

Where $\textbf{K}$ is the instruments matrix corresponding to the endogenous variables and  $y_{i,t-s}$ denominates $NIM_{i,t-k}$ for  Equation-\ref{eq6}.

Nevertheless, the first estimator is not free from bias and imprecision. Hence, in order to alleviate the possible bias and imprecision, as \cite{blundell1998initial} mentioned that a new estimator that unites a system in the difference estimator can be used if the regressors have limited time period that is known as “the Blundell and Bond system GMM”.  The econometric presumption is that the difference in the explanatory variables and the bank specific effect are uncorrelated. Thus, the stationary features are; 

For Equation-\ref{eq3};

\begin{equation}
  \mbox{\large E}\big[NIM_{i,t+p} \mu_{i}\big]=\mbox{\large E}\big[NIM_{i,t+q} \mu_{i} \big] \quad and \quad \mbox{\large E}\big[\textbf{X}_{i,t+p}\mu_{i}\big]=\mbox{\large E}\big[\textbf{X}_{i,t+q} \mu_{i} \big] \quad \forall \quad p\quad  and \quad q  
\end{equation}

The additional moment conditions;

\begin{equation}
  \mbox{\large E}\big[\Delta NIM_{i,t-k} (\mu_{i}+\varepsilon_{i,t})\big]=0 \quad for \quad k=1 
\end{equation}

\begin{equation}
  \mbox{\large E}\big[\Delta \textbf{X}_{i,t-k} (\mu_{i}+\varepsilon_{i,t})\big]=0 \quad for \quad k=1 
\end{equation}

Now, the GMM methods can be used for model in order to estimate the consistent and efficient parameter by putting account the moment condition for Equation (6), (7), (9) and (10) for the determination of the NIM model. The system GMM employs the lagged dependent variable in the levels and in differences; at the same time other lagged regressors can be suffered from endogeneity problem \citep{dietrich2011determinants}. 

Finally, to control the health of estimation method this study performs some tests. In order to reject the null hypothesis of joint insignificance coefficients this study uses the Wald test. For the validity of the instrument in the system GMM this study applied two specification tests. First is the Sargan test which is used to for the emphasising over identifying restrictions is valid. Second is the Arellano-Bond test which is to investigate the hypothesis that residual term is serially uncorrelated.

\subsection{\textbf{Explanatory Variables}}

\subsubsection{Pure Interest Margin Variables }

\textbf{Risk Aversion:} The  ratio of equity to the total assets  as a proxy for the bank risk aversion or bank-capitalisation ratio. For instance, high-capitalised banks are generally thought to be safer and less risky than lower capitalised banks with higher interest rates for credits. Therefore, it is expected that the risk aversion has a positive impact on the NIM. On the other hand, \citep{brock2000understanding} argued that there is a negative correlation between the NIM and the risk aversion. Because the less capitalised banks have more incentives to take more risks, the consequence is higher margin in order to obtain higher return. As a result the impact of risk aversion is not clear

\textbf{Credit Risk:} This study calculates the credit risk or the ratio of bad debt using the ratio of non-performing loan to the total loan. It is believed that if this ratio rises due to the health of the bank assets, it will deteriorate the banks, and will generally raise their interest rate to compensate this cost. A positive relationship is expected between the NIM and ratio of bad debt.

\textbf{Operating Cost:} Operating costs are simply defined as the ratio of operating expenses to the total bank assets. \cite{demirgucc1999determinants} claimed that banks with high operating cost are willing to pass this cost to their customers. Therefore, it is clear that banks experiencing high operating cost are predicted to have high interest margins; hence, operating cost has a positive effect on the NIM.

\textbf{Bank Size:} Bank size is captured by the logarithm of bank`s total assets. Ex-ante, the relationship between the bank NIM and bank size is ambiguous. The general perception is that the governments are not willing to permit large banks to fail, -too big to fail-for this reason big banks might take a position that has high-risk but high returns. Hence,  the sign of the relationship between bank size and the NIM is predicted to be positive. On the other hand, some studies (for example \cite{demirgucc1999determinants}; \cite{laeven2007there} argue that big banks generally apply lower interest margins relatively to the smaller ones because of the scale efficiencies.

\subsubsection{Bank Specific Variables}

\textbf{Liquidity Ratio:} Liquidity ratio is proxied in relation to the liquid assets to total assets. The characteristics of the liquid assets tend to yield lower return \citep{aysan2010macroeconomic}. For this reason, the banks have high amount of liquid assets that are more likely to have less interest income. Thus, the predicted sign of liquid assets is negative.

\textbf{Efficiency-Management Quality:} Management quality is defined as the operating expense to total revenues. This relation is also used to measure the impact of management quality on the bank profitability. The operating expense is accepted as a necessary cost to create unit gross revenue. Therefore, the banks with high management efficiency are able to create and invest in high profitable assets. Hence, an increasing ratio means decreasing management efficiency, thus, a lower NIM, and as a natural result, the relationship between management efficiency and banks NIM is negative.

\textbf{Implicit interest payment:} Implicit interest payment is expressed as the difference between the non-interest cost and other operating revenue over total assets. The sign of the implicit interest payment is not clear. While \cite{maudos2004factors}; and \cite{maudos2009determinants} have found a positive impact,\cite{liebeg2006determinants}  and \cite{gounder2012determinants} have found a negative relationship.

\textbf{Deposit Growth Rate:} The deposit growth rate is measured by the quarterly growth of bank deposits. It is expected that the banks with high growth rate of deposit are able to decrease its NIM because of the economics of scale. Deposits growth rate depends on many different factors such as the number of branches and management quality. A negative relationship associated with deposit growth rate and the NIM is expected. 

\textbf{Operation Diversity:} The ratio of non-interest revenue to total operating income captures the operating diversity. This proportion suggests the information of  non-traditional banking activities. If this ratio is high for a bank, this means that that bank focuses the non-conventional banks operation such as fee based activities. This is important especially during the crises and uncertainty. This variable is used by many other studies. For example, \cite{lin2012determinants} for Asian banks, \cite{valverde2007determinants} for European banks, \cite{liebeg2006determinants} for Austrian banks. These operations are known to be less risky than interest-based operations, thus the interest return operations are of high risk but have high returns, vice versa. For instance, \cite{demirgucc1999determinants} showed that non-interest assets have less returns than the interest based assets. In order to diversify the operation, banks need a wide network and high-qualified employees and also bear some other expenses. Thus, the expectation is that operation diversity has a negative relationship with the NIM.

\begin{table}[H]
  \centering
  \caption{~ Definition of Variables and Their Expected Effect on the Net Interest Margin}
  \resizebox{\textwidth}{!}{  \begin{tabular}{l|llc}
  Variables & Notation & Definition & Expected Sign \\
    \midrule
    \midrule
    Net Interest Margin \% & NIM   & Net interest income divided by total assets &  \\
    Risk Aversion \% & RA    & Equity over total assets & \textbf{?} \\
    Credit Risk \% & RBD   & Non-performing loan over total loan & \textbf{+} \\
    Operating Cost \% & OC    & Operation cost over total assets & \textbf{?} \\
    Bank Size  & LOGTA & Logarithm of total assets & \textbf{?} \\
    Liquidity Ratio \% & LQR   & Ratio of liquid assets to total assets & \textbf{-} \\
    Management Quality \% & MNGMT & Total expenses over total  generated revenues & \textbf{-} \\
    Implicit Interest Payment \% & IIP   & Net non-interest income over total assets & \textbf{+} \\
    Deposits Growth \% & DPZTG & Quarterly growth of deposits & \textbf{-} \\
    Operation Diversity \% & DVRSTY & Non-interest income over operating income & \textbf{-} \\
    Herfindahl Index \% & HHI   & Herfindahl index for assets & \textbf{+} \\
    Real GDP Growth \% & GDP   & Quarterly real GDP growth & \textbf{?} \\
    Inflation \% & INF   & CPI growth rate & \textbf{+} \\
    \bottomrule
    \bottomrule
    \end{tabular}}%
  \label{tab_define}%
\end{table}%

\subsubsection{Macroeconomic and Market Variables}

\textbf{Competitive Structure:} To capture the competitive structure of the banking industry, the Herfindahl–Hirschman Index (HHI) is used. The HHI is defined as the sum of the squares of the market share of the individual bank assets in the total banking assets in a given time (in this case: quarterly). It is generally accepted that the high market concentration reflects less competition and enables banks to have monopolistic power over the interest rates. Therefore, most studies expect the sign of the estimated coefficient of the HHI to have a positive sign. On contrary, some studies (for example, see \cite{barajas2000impact} argue that the NIM and market concentration have a negative relationship. For example, their evidence shows that a higher bank concentration could be the consequence of a strong competition in the banking market, which would offer an opposite relationship.  As a result, the overall impact of industry concentration on the NIM is not clear and still waits to be answered empirically. 

\textbf{The Real GDP Growth:} To measure the effect of the business cycle on the NIM, this study controls for the real GDP growth. The impact of the GDP growth on the NIM varies over countries; therefore the expected sign is not clear. While Khawaja and Din (2007) have found a negative relationship between the GDP growth and the NIM in Pakistan, \cite{liebeg2006determinants} have suggested a positive relationship for Austrian banking sector. In parallel with these findings, \cite{costa2007macroeconomic} suggest that the relationship is ex-ante, ambiguous. 

\textbf{Inflation:} As a macroeconomic uncertainty indicator, this study uses inflation variable to measure the impact of macroeconomic uncertainty on the NIM. Due to difficulty of anticipating the inflation rate for the next period, banks generally prefer to hold a safe position such as investing in government bonds instead of lending loan. Because an unpredicted inflation rate may raise costs, there is a reason for imperfect interest rate adjustment. Therefore, in a high volatile economic environment, banks might charge higher interest margin for lending to cover the potential risk and this study expects a positive relationship between inflation and the NIM. 

\section{Empirical Results}
\subsection{\textbf{Overall Results}} 
This section has been divided into two parts. The first part provides the findings of the whole sample and the second part shows the results of the separate estimations by bank ownership structure. To investigate the hypotheses, this study estimates four different models with using proper econometric tests to decide appropriate estimation technique. Table-\ref{tab_result} summaries the regression results employing different techniques. The column (1) shows the Pooled OLS (POLS) result, since, the POLS does not allow to accounting unobserved bank specific effect, the within Fixed Effect (FE) and GLS-Random Effect(RE) methods are executed. The column (2) and (3) provide these results respectively. The results of RE estimation are consistent with the results of the POLS. To investigate the relevance of bank specific effects, the LM test provides that this study rejects the null hypothesis, which is the POLS is a proper method to provide the relationship between the NIM and its determinants. This means that the FE or the RE should be used instead of the POLS in cases of static estimation methods. To decide the Fixed or Random Effect, the Hausman test technique is used and the result which is in favour of FE Model. All these tests results are summarise in Table-\ref{tab_result}.

\begin{table}[htbp]
  \centering
  \caption{~Regressions Results}
    \begin{tabular}{p{9.915em}rrrr}
    \toprule
    & \textbf{1} & \textbf{2} & \textbf{3} & \textbf{4} \\
    \textbf{VARIABLES} & \textbf{POLS} & \textbf{FE} & \textbf{RE} & \textbf{GMM} \\
    \midrule
    L.NIM & - & - & - & 0.228*** \\
    &       &       &       & (0.0258) \\
    L2.NIM & - & - & - & -0.0122*** \\
    &       &       &       & (0.00223) \\
    RA    & 0.0899*** & 0.0983*** & 0.0899*** & 0.0115 \\
    & (0.0127) & (0.0156) & (0.0127) & (0.00777) \\
    RBD   & -0.0164** & -0.0218*** & -0.0164** & -0.0343*** \\
    & (0.0069) & (0.00775) & (0.0069) & (0.00524) \\
    OC    & 0.134** & 0.153** & 0.134** & 0.0817*** \\
    & (0.0561) & (0.06) & (0.0561) & (0.0107) \\
    LOGTA & 0.413*** & 0.834 & 0.413*** & 0.00962 \\
    & (0.12) & (0.509) & (0.12) & (0.147) \\
    LQR   & -5.98E-05 & -0.00985* & -5.98E-05 & -0.002 \\
    & (0.00417) & (0.00563) & (0.00417) & (0.00176) \\
    MNGMT & -0.00381*** & -0.00344*** & -0.00381*** & -0.00620*** \\
    & (0.000885) & (0.000898) & (0.000885) & (0.00023) \\
    IIP   & 0.508*** & 0.541*** & 0.508*** & 0.369*** \\
    & (0.0198) & (0.0209) & (0.0198) & (0.0138) \\
    DPZTG & -0.000503*** & -0.000536*** & -0.000503*** & -0.000344*** \\
    & (0.000166) & (0.000165) & (0.000166) & (1.71E-05) \\
    DVRSTY & -0.228** & -0.304*** & -0.228** & -0.379*** \\
    & (0.0933) & (0.0972) & (0.0933) & (0.0489) \\
    HHI   & 0.0528 & 0.181 & 0.0528 & 0.221 \\
    & (0.182) & (0.255) & (0.182) & (0.151) \\
    GDP   & -0.0294** & -0.0262* & -0.0294** & -0.00215 \\
    & (0.0148) & (0.0147) & (0.0148) & (0.00205) \\
    INF   & 0.0394*** & 0.0445*** & 0.0394*** & 0.0295*** \\
    & (0.00819) & (0.00842) & (0.00819) & (0.00599) \\
    CONSTANT & -3.546 & -7.657 & -3.546 & -1.354 \\
    & (2.25) & (5.83) & (2.25) & (2.261) \\
    \midrule
    OBSERVATIONS & 966   & 966   & 966   & 920 \\
    R-squared & 0.535 & 0.538 & 0.535 & 0.555 \\
    Sargan test (P value) & - & - & - & 0.6743 \\
    A-Bond Test AR(1) & - & - & - & 0.0226 \\
    A-Bond Test AR(2) & - & - & - & 0.1157 \\
    Number of Banks & 23    & 23    & 23    & 23 \\
    \bottomrule
    \end{tabular}%
  \label{tab_result}%
\end{table}%

Nevertheless, such static models do not allow us to investigate the potential dynamism, thus, performing the dynamic GMM estimator in this regard seems a best alternative, and the column (4) shows the results of the GMM estimation. This hypothesis considers that the lagged value of the NIMs might have a significant effect on the current value of the NIM. As it can be seen from the first and the second rows` result in column (4) on the Table-\ref{tab_define}, the lagged dependent variables have a significant effect on the current values of the NIM. Additionally, we also provide the Sargan test for over-identifying restrictions and the results prove that our specification is well modelled. Furthermore, the results of the Arellano-Bond test for checking serial correlation support to our model. The following results of variables are based on the baseline specification which uses the GMM estimation method.

\textbf{Risk Aversion:} On the contrary of a number of studies (see \cite{claeys2008determinants}; \cite{maudos2009determinants}; \cite{flamini2009determinants}; \cite{fungavcova2011determinants}) the results of this study surprisingly have shown that the correlation between risk version and the NIM is insignificant even its coefficient is positive. Higher risk aversion ratio implies that banks set higher margin due to positive relationship.

\textbf{Credit Risk:} Surprisingly, the relationship between credit risk and the NIM is not positive as predicted. However, it has a significant impact on the NIM. A positive and significant relationship implies that the NIM decreases as the quality of credit falls and the banks with large credit risk might raise margin in order to solve such problems. The result is inconsistent with \cite{gounder2012determinants}

\textbf{Operating Cost:} The coefficient of operating costs is positive and statistically significant. This means that the banks with higher operating expenses have higher NIM to compensate their operation expense. Hence, high operating costs are mostly passed to customers to keep the banks` profit unaffected. The result of the operating cost is in line with \cite{maudos2009determinants}

\textbf{Bank Size:}  The bank size does not seem to be a significant determinant of banks` NIM and have a negative sign. This study also used the banks` size variable for the logarithm of total loan, instead of the logarithm of total assets, but again failed to find any significant relationship between the bank size and NIM. Even the coefficient of the bank size is insignificant, the sign is negative and it means that big banks are assumed to set lower interest margin. This result is, again, inconsistent with many studies focusing on developing countries such as \cite{tan2012determinants}.

\textbf{Liquidity Ratio:}  Results show that there is a negative relationship between liquidity ratio and the NIM. However, the magnitude of the impact is insignificant. This finding is in line with \cite{hawtrey2008bank} and \cite{maudos2009determinants}.

\textbf{Management Quality:}  This study`s result suggests that the management efficiency a negative and significant effect on the margin. This result implies that the banks with less management quality set higher interest margin. This relation can be interpreted as a beneficial condition for the bank`s client that higher management quality encourages banks to exhibit higher deposit rates and lower loan rates. This study`s result is consistent with \cite{liebeg2006determinants}, \cite{hawtrey2008bank}, \cite{claeys2008determinants} \cite{horvath2009determinants}.

\textbf{Implicit Interest Payment:} Result has suggested that there is a statistically significant and positive relationship between the interest margin and the implicit interest payment. This relation implies that banks in Turkey might try to recover the implicit interest payment via margin setting \citep{gounder2012determinants}. Hence, the banks that set their services more implicitly through less compensation of liabilities exhibit a higher margin \citep{maudos2009determinants}. This finding is in line with \cite{saunders2000determinants},

\textbf{Deposits Growth:}  A significant and negatively relationship between the deposit growth and the NIM has been found. It means that  the banks with the ability of collecting deposits of high rate exhibit lower interest margin.

\textbf{Operation Diversity:}  The result suggests that the operation diversity and the NIM have a significant and a negative relationship. The result suggest that the banks engaging mostly in interest related operations, in other words the ones who take more risks and who are less diverse exhibit greater interest margins, vice versa. As a result, if a bank takes high risks, it gains more returns than a bank that takes less risks.

\textbf{Competitive Structure (Herfindahl Hirschman Index):}  The coefficient of the Herfindahl-Hirchman Index is positive but insignificant. Hence, this result indicates that, ceteris paribus, the banks with high market do not exploit their market power on determining the interest margin in Turkey. This result is consistent with \cite{flamini2009determinants}, and \cite{fungavcova2011determinants}. 
The Real GDP Growth: This study, surprisingly, has not found any significant correlation between the real economic growth and the NIM.. Although the coefficient of the GDP is insignificant, its sign is negative which means that economic growth might keep interest margin low.  This study`s result is in line with \cite{claessens2001does}.

\textbf{Inflation:} As expected the coefficient of inflation is positive and it affects the NIM significantly, which means that banks estimate the future movement in inflation accurately and hastily enough to adjust rates and interest margin \citep{flamini2009determinants}. This result can be explained with the mathematical expression. Assuming that $\gamma_{D}$ and $\gamma_{L}$ are the real interest rate on deposits and loan, respectively, as that the fisher equation holds, bank interest margin can be expressed in nominal value as:

\begin{equation}
 (1+\gamma_{L}) (1+\pi)-(1+\gamma_{D})(1+\pi)
\end{equation}

This after manipulation gives: 

\begin{equation}
(\gamma_{L}-\gamma_{D})(1+\pi)\end{equation}

Where, $\pi$ indicates the inflation rate. Therefore, the impacts of inflation on the nominal interest rates, deposits and loans do not cancel out for the reason of the cross product term, indicating a positive impact of inflation on the NIM \citep{flamini2009determinants}. Hence, our finding implies that banks adjust a higher interest margin in a higher inflation condition, vice versa. Also our finding is consistent \cite{demirgucc1999determinants} and \cite{maudos2004factors}.

\subsection{\textbf{Ownership Results}}

The possible effects of ownership on the NIM are investigated in this section. To obtain a clear and robust result, this study has subdivided sample into three parts according to ownership, foreign, state and private banks respectively, to analyse variations on the effects of the NIM determinants across ownership structure. Also this can be seen on Chow Test results, which are provided by Table-10, Table-11 and Table-12 on the appendix. According to Chow test results the coefficients do not have the same affect for  the three ownership groups. For example, for foreign banks the rejection of the hypothesis (p-value <0.05) means that the foreign banks DO NOT share the same the coefficient for the corresponding variable.

Table-\ref{tab_owner} summarises the findings for ten foreign banks, three state banks and sixteen private banks by comparing the main estimated results, which is in the column (4). 

At the first glance, the results of private-owned banks are very similar to the main sample results in terms of both coefficients` signs and their effects on the NIM. All coefficients` of private banks and main sample sign are the same except the sign of risk aversion. This result verifies our hypothesis that the structure of the Turkish banking system is largely controlled by domestic private banks. Another common result is that the coefficients of the operating cost, management quality implicit interest payment and operation diversity variables and their significant effects on the NIM are consistent across all ownership groups. This result implies that all banks react similarly to changes in management quality, implicit interest payment ratio, operating cost and operation diversity when determining the NIM. Although the management quality has a significant effect on the NIM, its economic impacts are very small for all banks. 
% Table generated by Excel2LaTeX from sheet 'Sheet7'
\begin{table}[htbp]
  \centering
  \caption{~Estimation Results by Bank Ownership}
    \begin{tabular}{lcccc}
    \toprule
          & \textbf{1} & \textbf{2} & \textbf{3} & \textbf{4} \\
   \textbf{VARIABLES} & \textbf{FOREIGN} & \textbf{STATE} & \textbf{PRIVATE} & \textbf{MAIN} \\
    \midrule
    L.NIM & 0.256 & 0.200*** & 0.115*** & 0.228*** \\
          & (0.176) & (0.0521) & (0.0429) & (0.0258) \\
    L2.NIM & -0.0294** & 0.243** & 0.0461 & -0.0122*** \\
          & (0.0138) & (0.124 ) & (0.0468) & (0.00223) \\
    RA & 0.0264 & -0.00692 & -0.0125 & 0.0115 \\
          & (0.0246 )& (0.0234) & (0.0166) & (0.00777) \\
    RBD & -0.089 & 0.0103*** & -0.0313*** & -0.0343*** \\
          & (0.125) & (0.00382) & (0.00549) & (0.00524) \\
    OC & 0.0213 & 0.182 & 0.0381 & 0.0817*** \\
          & (0.115) & (0.173) & (0.25) & (0.0107) \\
    LOGTA & 0.417** & -0.17 & -0.094 & 0.00962 \\
          & (0.175 )& (0.193) & (-0.175) & (0.147) \\
    LQR & -0.00827** & 0.00276*** & -0.00656*** & -0.002 \\
          & (0.004) & (0.000941) & (0.00138 )& (0.00176) \\
    MNGMT & -0.00617*** & -0.00570** & -0.00780*** & -0.00620*** \\
          & (0.00157) & (0.00241) & (0.00235) & (0.00023 )\\
    IIP & 0.376*** & 0.0308 & 0.425*** & 0.369*** \\
          & (0.0555) & (0.104 )& (0.0623) & (0.0138) \\
    DPZTG & -0.000389*** & 0.0027 & -0.000549 & -0.000344*** \\
          & (0.000146) & (0.00244) & (0.00268) & (1.71E-05) \\
    DVRSTY & -0.199** & -1.125*** & -0.593*** & -0.379*** \\
          & (0.0998 )& (0.216 )& (0.181 )& (0.0489) \\
    HHI & -0.0221 & -0.0568*** & 0.0692* & 0.221 \\
          & (0.446) & (0.0135) & (0.0412) & (0.151) \\
    GDP & 0.0168 & -0.00392 & -0.00844*** & -0.00215 \\
          & (0.0232) & (0.00254) & (0.0027) & (0.00205) \\
    INF & 0.0847** & -0.00137 & 0.00672 & 0.0295*** \\
          & (0.0418) & (0.00711) & (0.00755) & (0.00599) \\
    CONSTANT & -1.708 & 2.878* & 1.939 & -1.354 \\
          & (4.475) & (1.717) & (1.625) & (2.261) \\
          &       &       &       &  \\
    \midrule
    Observations & 288   & 120   & 498   & 920 \\
    Number of Banks & 10    & 3     & 16    & 23 \\
    Sargan test (P value) & 0.3212 & 0.0278 & 0.4563 & 0.6743 \\
   A-Bond Test AR(1) & 0.1284 & 0.0872 & 0.0062 & 0.0226 \\
    A-Bond Test AR(2) & 0.2492 & 0.1783 & 0.2634 & 0.1157 \\
          &       &       &       &  \\
    \bottomrule
    \end{tabular}%
  \label{tab_owner}%
\end{table}%

Regarding the foreign ownership, the results have showed that the foreign banks are not similar to state and private banks in two aspects. The first distinctive feature of foreign banks is that the banks size positively and significantly affects only for the NIM of foreign banks. The positive and significant coefficient for the bank size implies that the big foreign banks set higher margins in the Turkish banking system. Second, the risk aversion (or capitalisation ratio) has a positive effect on only for the NIM of foreign banks. This indicates that foreign banks with higher capitalisation ratio tend to set higher NIM.

Considering the private domestic banks, this study has found an interesting result for private banks. Solely, the NIM of private banks is positively affected by alteration in market structure. The positive coefficient and significant effect for the market structure indicate that the private domestic banks exploit their special position in the industry by setting higher NIMs. 

In terms of state banks, this study has also found some important results. One of them is that the liquidity ratio is a positive determinant of NIM only for state banks. In contrast to this study`s expectation, the effect of liquidity risk on NIM is positive for foreign banks. Another important result is that the effect of the rate of bad debt or credit risk. Credit risk has a negative impact on the NIM of foreign and private banks and a positive effect on for only state bank. The negative sign indicates a fierce competition on gaining the market share and thus, the results have showed that foreign and private banks are more willing to accept higher ratio of bad debt without increasing their interest margins for the sake of obtaining more market share in the sector.

Consequently, the results of the second estimation have showed that considerably variations exist in terms of the effect of the NIM factors across the ownership groups. Thus, it is crucial to take into the ownership structure account while investigating the effect of the interest margin determinants in Turkey. Otherwise, a possible disregarding may cause inaccurate conclusions.

\section{Robustness Checks}

This section analyses the robustness and sensitivity of our findings using six different robustness checks. A set of robustness tests is reported in Table-\ref{tab_robust}. Firstly, this study dropped three banks from the main sample and re-estimated by employing the same variables and the same techniques. These banks are from different bank-ownership groups. Secondly, we dropped the whole state-owned banks from the sample and re-estimated the model same as the previous methods. Thirdly; we have re-estimated the model for only domestic private and state banks by excluding foreign banks from the whole sample. Fourthly, we employ an alternative measure of bank size. In the foundation model, logarithm of total assets was considered as the variable of bank size. However, this time, the bank size is measured by market share as \cite{liebeg2006determinants} used in their paper as a determinant of the NIM.   Later, this study added a bank-specific variable which is the credit size as a determinant of the NIM (see, \cite{aysan2010macroeconomic}). Lastly, we added a macro-specific determinant of the NIM, which is the interbank interest rate. 

Using different sample size and types this study has re-estimated the model with the same explanatory variables for column (1), (2) and (3). As a result, the sign of coefficients of the explanatory variables, except for the RA, LOGT and HHI (in only one case) are remarkably consistent over different sample size. Also their significances are very similar. Therefore, these three different samples specifications support that the results obtained for the baseline model are valid. 

In addition, by employing an alternative bank size variable we re-estimated the model by using the same techniques and such results are reported in column (4). The last robustness check`s results are also consistent with the main results.

Finally, in the fifth and sixth robustness checks by including a bank-specific and a macro specific variable, respectively, into the model is also in favour of the validity of the results of this study. The results in the columns (5) and (6) are considerably parallel to each other such as all coefficients` sign are the same and their significant effects have too small variation with no exception. Consequently, all robustness tests using different variables and different sample size support the baseline estimation results. The coefficient sign of the explanatory variables and their magnitude on the NIM, and the main results also are in line with each other.

% Table generated by Excel2LaTeX from sheet 'Sheet7'

\section{Conclusion}

This research has analysed how the pure-specific and bank-specific characteristics, and also macro and market-specific factors affect the NIM for almost all of the commercial banks in Turkey over the period from the last quarter of 2001 to the first quarter of 2012 with a particular emphasis on the role of the bank ownership by employing micro and macro level data.

Findings of this study clearly indicate that the NIM of a bank is mainly determined by the bank-specific characteristics such as management quality, operating cost, ratio of bad debt (credit risk), implicit interest payment, bank`s deposit growth rate and operation diversity, and also inflation as a macro-specific factor. Regarding management efficiency, this study finds that efficient banks exhibit lower interest margin and charge lower fees in favour of costumers. This result supports the hypothesis that management quality can improve the financial intermediation system. Also, findings of this study suggest that implicit interest payment causes a higher interest margin since this determinant represents an extra cost for the banks. Furthermore, our research has found that ratio of bad debt (or credit risk) has a significant and negative impact on the bank margin. Thus, the banks with high credit risk level exhibit lower margin. Another important point is that high operating cost raises the interest margin since the banks with high cost may pass these expenses on to their clients by charging higher rates of interest on loan and providing lower rates for deposits.

Considering the external drivers related to the macroeconomic environment variables such as the real economic growth and the price stability on the determination of the NIM, we have found a strong and positive relationship between inflation and the NIM. It can be interpreted that high inflation rate contributes to a higher margin, thus; it has a deterioration impact on the financial intermediation system. In contrast, this study failed to find any significant relation between the GDP growth and the NIM.

The ownership-related findings have supported the hypothesis that the bank ownership has a strong impact on the determination of the NIM. Thus, bank ownership has a crucial role on the determination of the NIM and should not be disregarded.

Overall, the results of this study have showed that the NIM is a crucial element in order to maintain financial stability of Turkey. Hence, this study has some policy recommendations for both bank managers and the government authorities. On the side of bank managers, they should upgrade their management quality and decrease the operating cost, since both are significant determinant of the NIM. Also they should investigate on the new technologies such as enhancing the ATM network and encourage their customers to use online banking for the sake of reducing implicit interest payment, which is another major determinant of the NIM. On the side of government authorities, the price stability is one of the main determinants of the NIM because the high inflation decreases loan expansion by causing higher interest margin. Thus, monetary policy should target to control inflation very strictly by keeping a reasonable rate, in order to foster the strong financial intermediation system in Turkey.

\clearpage

\appendix

\section{~Appendix}
\label{sec:app1}

\begin{table}[htbp]
  \centering
  \caption{~ Breusch and Pagan LM Test}
    \begin{tabular}{lrrrr}
    \toprule
    \multicolumn{5}{c}{\textbf{Breusch  and Pagan Lagrangian multiplier test for random effects}} \\
    \midrule
    \midrule
    \multicolumn{5}{c}{nim[bank,t] = Xb + u[bank] + e[bank,t]} \\
    \midrule
    \midrule
    Estimated results: &       &       &       &  \\
    \midrule
    \midrule
          & \multicolumn{1}{l}{Var} & \multicolumn{1}{l}{ $sd = \sqrt{Var}$} &       &  \\
\cmidrule{1-3}          &       &       &       &  \\
    \multicolumn{1}{r|}{nim} & 12.22123 & 3.495887 &       &  \\
    \multicolumn{1}{r|}{e} & 5.688468 & 2.385051 &       &  \\
    \multicolumn{1}{r|}{u} & 0     & 0     &       &  \\
    Test:   Var(u) $=$  0 &       &       &       &  \\
    chibar2(01)$=$   0.00 &       &       &       &  \\
    Prob $> \chi^2 =   1.0000$ &       &       &       &  \\
    \bottomrule
    \end{tabular}%
  \label{tabbplm}%
\end{table}%

% Table generated by Excel2LaTeX from sheet 'Sheet3'
% Table generated by Excel2LaTeX from sheet 'Sheet3'
\begin{table}[htbp]
  \centering
  \caption{~Hausman Test}
    \begin{tabular}{ccccr}
    \toprule
      \midrule
   & \multicolumn{2}{c}{ ---- Coefficients ----} &       &  \\
    & \multicolumn{1}{l}{(b)} & \multicolumn{1}{c}{(B)} & \multicolumn{1}{c}{(b-B)} & \multicolumn{1}{c}{$\sqrt{(diag (V_{b} - V_{B})}$} \\
    \hline
& \multicolumn{1}{l}{FE} & \multicolumn{1}{c}{RE} & \multicolumn{1}{c}{Difference} & \multicolumn{1}{c}{S.E.} \\
   
    \midrule
    \multicolumn{1}{l|}{RA} & 0.098 & 0.090 & 0.008 & \multicolumn{1}{c}{0.009} \\
    \multicolumn{1}{l|}{RBD} & -0.022 & -0.016 & -0.005 & \multicolumn{1}{c}{0.004} \\
    \multicolumn{1}{l|}{OC} & 0.153 & 0.134 & 0.019 & \multicolumn{1}{c}{0.021} \\
    \multicolumn{1}{l|}{LOGTA} & 0.834 & 0.413 & 0.422 & \multicolumn{1}{c}{0.495} \\
    \multicolumn{1}{l|}{LQR} & -0.010 & 0.000 & -0.010 & \multicolumn{1}{c}{0.004} \\
    \multicolumn{1}{l|}{MNGMT} & -0.003 & -0.004 & 0.000 & \multicolumn{1}{c}{0.000} \\
    \multicolumn{1}{l|}{IIP} & 0.541 & 0.508 & 0.034 & \multicolumn{1}{c}{0.007} \\
    \multicolumn{1}{l|}{DPZTG} & -0.001 & -0.001 & 0.000 & \multicolumn{1}{c}{.} \\
    \multicolumn{1}{l|}{DVRSTY} & 0.304 & 0.228 & 0.076 & \multicolumn{1}{c}{0.027} \\
    \multicolumn{1}{l|}{HHI} & 0.181 & 0.053 & 0.128 & \multicolumn{1}{c}{0.178} \\
    \multicolumn{1}{l|}{GDP} & -0.026 & -0.029 & 0.003 & \multicolumn{1}{c}{.} \\
    \multicolumn{1}{l|}{INF} & 0.045 & 0.039 & 0.005 & \multicolumn{1}{c}{0.002} \\
    \midrule
    \midrule
    \multicolumn{5}{l}{b = consistent under Ho and Ha; obtained from xtreg}    \\
   \multicolumn{5}{l}{B = inconsistent under Ha, efficient under Ho; obtained from xtreg}     \\
            \\
   \multicolumn{5}{l}{ Test:  Ho:  difference in coefficients not systematic}    \\
            \\
    \multicolumn{5}{l}{$ \chi^{2}(11)=(b-B) [(V_{b} - V_{B})^{(-1)}](b-B) $ }    \\
    \multicolumn{5}{l}{$=$  333.71 }      \\
 \multicolumn{5}{l}{    Prob 	$> \chi^{2} =$     0.0000}     \\
 \multicolumn{5}{l}{$(V_{b}- V_{B}$ is not positive definite)}         \\

    \end{tabular}%
  \label{tabhaus}%
\end{table}%

\begin{table}[H]
  \centering
  \caption{~Robustness Tests}
  \resizebox{\textwidth}{!}{   \begin{tabular}{lcccccc}
    \toprule
               & 1    & 2    & 3    & 4    & 5    & 6 \\
    VARIABLES & SSMPL & NOSTT & NOFRGN & NEWSIZE & CRDT & IIRATE \\
    \midrule
    L.NIM & 0.138*** & 0.216** & 0.118*** & 0.207*** & 0.209*** & 0.209*** \\
    & (0.034) & (0.0838) & (0.0409) & (0.0675) & (0.0684) & (0.069) \\
    L2.NIM & 0.033 & -0.014 & 0.0361 & -0.0172 & -0.0162 & -0.0158 \\
    & (0.04) & (0.0108) & (0.0422) & (0.0128 )& (0.0126) & (0.0134) \\
    RA    & -0.00245 & 0.0137 & 0.000158 & 0.0106 & 0.0119 & 0.0125 \\
    & (0.0111) & (0.0144) & (0.0141) & (0.0139) & (0.0136) & (0.0139) \\
    RBD   & -0.0227*** & -0.0559 & -0.0274** & -0.0424 & -0.0422 & -0.042 \\
    & (0.00814) & (0.0399) & (0.011 )& (0.0288) & (0.029) & (0.0281) \\
    OC    & 0.109 & 0.0905 & 0.0492 & 0.111 & 0.106 & 0.105 \\
    & (0.175) & (0.0738) & (0.231) & (0.073) & (0.0728) & (0.0735) \\
    LOGTA & -0.181* & 0.0731 & -0.162 & ------ & 0.0109 & 0.0263 \\
    & (0.108) & (0.178) & (0.144) & ------ & (0.142) & (0.201) \\
    LQR   & -0.00436** & -0.00313 & -0.00511*** & -0.0035 & -0.00323 & -0.00318 \\
    & (0.00185) & (0.00254) & (0.00132) & (0.00232) & (0.0029) & (0.00259) \\
    MNGMT & -0.00846*** & -0.00566*** & -0.00843*** & -0.00590*** & -0.00590*** & -0.00592*** \\
    & (0.00206) & (0.00161) & (0.00243) & (0.00168) & (0.00169) & (0.00161) \\
    IIP   & 0.393*** & 0.377*** & 0.405*** & 0.377*** & 0.377*** & 0.377*** \\
    & (0.0494) & (0.0459) & (0.0574) & (0.0433) & (0.0435) & (0.044) \\
    DPZTG & -0.000154 & -0.000326*** & 0.000786 & -0.000328*** & -0.000323*** & -0.000322*** \\
    & (0.000815) & (7.15E-05) & (0.00232) & (7.20E-05) & (6.72E-05) & (6.28E-05) \\
    DVRSTY & -0.627*** & -0.355*** & -0.632*** & -0.368** & -0.363** & -0.364** \\
    & (0.162) & (0.138) & (0.19) & (0.144) & (0.148) & (0.143) \\
    HHI   & 0.0712 & -0.0416 & 0.0544 & -0.0525 & -0.0572 & -0.0585 \\
    & (0.0468) & (0.213) & (0.0433) & (0.213) & (0.193) & (0.168) \\
    GDP   & -0.00621*** & -0.000954 & -0.00596** & -0.00141 & -0.00093 & -0.000753 \\
    & (0.00206) & (0.00776) & (0.00265) & (0.00681) & (0.00687) & (0.0076) \\
    INF   & 0.00695 & 0.0336 & 0.00888 & 0.0324 & 0.0325 & 0.0304 \\
    & (0.00582) & (0.0212) & (0.00693) & (0.0207) & (0.0216) & (0.0198) \\
    MS & ------ & ------ & ------ & -0.00789 & ------ & ------ \\
    &       &       &       & (0.0315) &       &  \\
    CRDT & ------ & ------ & ------ & ------ & 0.0161 & ------ \\
    &       &       &       &       &(0.484) &  \\
    IIR   & ------ & ------ & ------ & ------ & ------ & 0.00189 \\
    &       &       &       &       &       & (0.00812) \\
    Constant & 2.288** & 1.041 & 2.402* & 1.744 & 1.647 & 1.543 \\
    & (1.112) & (2.18) & (1.37) & (2.248) & (2.119) & (2.093) \\
    Observations & 800   & 800   & 618   & 920   & 920   & 920 \\
    Number of bank & 20    & 20    & 19    & 23    & 23    & 23 \\
    \bottomrule
    \end{tabular}}
  \label{tab_robust}%
\end{table}%

\begin{landscape}
\begin{table}[htbp]
  \centering
  \caption{~Selected Literatures for the NIM- Developed Countries}
    \begin{tabular}{lcccccc}
    \toprule
    Authors & Angbazo & \makecell{Saunders \\ \& \\ Schumacher}& \makecell{Maudos \\ \& \\ Fernandez de Guevara} & \makecell{Liebeg \\ \& \\ Schwaiger} & \makecell{Carbo Valverde\\ \& \\ Rodriguez Fernandez} & Williams \\
    \midrule
    \midrule
    Year & 1997  & 2000 & 2004  & 2006  & 2007  & 2007 \\
   Journal & JBF   & JIMF & JBF   & OeNB  & JBF   & FMII \\
    \midrule
    \midrule
    Risk Aversion & +     & + & +     & +     & +     & + \\
    Credit Risk & +     & N/A & +     & -     & +     & - \\
    Operating Cost & NA    & N/A & +     & +     & +     & + \\
    Bank Size & NA    & N/A & -     & -     & NA    & x \\
    Liquidity Ratio & -     & + & +     & NA    & NA    & x \\
    Management Quality & -     & N/A & -     & -     & NA    & - \\
    Implicit Interest Payment & +     & N/A & +     & -     & NA    & + \\
    Deposits Growth & NA    & N/A & NA    & NA    & NA    & NA \\
    Operation Diversity & NA    & N/A & NA    & NA    & NA    & NA \\
    Market Concentration & +     & + & +     & +     & +     & + \\
    Real GDP Growth & NA    & N/A & NA    & +     & -     & NA \\
    Inflation & NA    & N/A & NA    & NA    & NA    & NA \\
    Ownership & NA    & N/A & NA    & NA    & NA    & Foreign(-) \\
    \midrule
    Sample & USA   & \makecell{Germany, Spain, \\ France, UK, USA \\ Italy, Switzerland, } & \makecell{France, Germany,\\ Italy, Spain, UK} & Austria & \makecell{Germany, Spain, \\ France, the Netherlands, \\ Italy, UK, Sweden } & Australia \\
    Estimation Methods & GLS   & \makecell{Cross-sectional \\ OLS} & \makecell{Fixed Effect \\ OLS } & Dynamic GMM & Dynamic GMM & \makecell{POLS \\ RE } \\
    \bottomrule
    \bottomrule
    \end{tabular}%
  \label{tab:addlabel}%
\end{table}%
\end{landscape}

\begin{landscape}
\begin{table}[htbp]
  \centering
  \caption{~Selected Literatures for the NIM- Developing Countries}
  \scalebox{0.85}{  \begin{tabular}{lccccccccc}
    \toprule
    Authors & Drakos & \makecell{Martinez Peria\\ \& \\ Mody } & \makecell{Claeys \\ \& \\ Vander Vennet } & \makecell{Schwaiger \\  \& Liebeg } & \makecell{Maudos \\ \& \\  Solisa } & Horvath & \makecell{Fungacova\\ \& \\ Poghosyan} & \makecell{Gounder\\ \& \\ Sharma } & Tan \\
    \midrule
    \midrule
    Year  & 2003  & 2004  & 2008  & 2008  & 2009  & 2009  & 2011  & 2012  & 2012 \\
    Journal & JPM   & JMCB  & ES    & OeNB  & JBF   & Czech JEF & ES    & APE   & IMF \\
    \midrule
    \midrule
Risk Aversion & +     & x     & +     & +     & +     & -     & +     & +     & + \\
   Credit Risk & +     & x     & NA    & +     & -     & +     & -     & +     & NA \\
   Operating Cost & NA    & NA    & NA    & +     & +     & +     & +     & +     & + \\
  Bank Size & NA    & NA    & -     & x     & +     & -     & -     & NA    & - \\
   Liquidity Ratio & -     & +     & NA    & NA    & +     & NA    & -     & -     & NA \\
  Management Quality & NA    & +     & -     & NA    & -     & -     & NA    & -     & NA \\
   Implicit Interest Payment & NA    & NA    & NA    & +     & +     & NA    & NA    & +     & NA \\
    Deposits Growth & NA    & NA    & NA    & +     & NA    & NA    & NA    & NA    & NA \\
    Operation Diversity & NA    & NA    & NA    & NA    & NA    & NA    & NA    & NA    & NA \\
    Market Concentration & NA    & +     & +     & +     & +     & x     & -     & +     & + \\
    Real GDP Growth & NA    & x     & +     & +     & NA    & x     & NA    & NA    & - \\
    Inflation & NA    & x     & +     & NA    & NA    & +     & NA    & NA    & + \\
    \midrule
    Ownership & Foreign (-) & Foreign (-) & NA & Foreign(+) State(x) & NA & NA & + & NA & Foreign(+) \\
    \midrule
    \midrule
    Sample & CEE Countries & Latin America & CEE Countries & CEE Countries & Mexico & Czech & Russia & Fiji & Philippines \\
    Estimation Methods & GLS   & Pooled OLS & RE OLS & FE OLS & FE/GMM & GMM   & FE OLS & RE OLS & FE OLS \\
    \bottomrule
    \bottomrule
    \end{tabular}}%
  \label{tab:addlabel}%
\end{table}%
\end{landscape}

\begin{table}[htbp]
  \centering
  \caption{~Selected Literatures for the NIM- Cross-Country Countries}
    \begin{tabular}{lccc}
    \toprule
    Authors & \makecell{ Demirguc-Kunt\\ \& \\  Huizinga} & \makecell{Hawtrey\\ \& \\ Liang } & Kasman et al. \\
    \midrule
    \midrule
    Year  & 1999  & \multicolumn{1}{c}{2008} & \multicolumn{1}{c}{2010} \\
    Journal & WB Econ Review & \multicolumn{1}{c}{NAJEF} & \multicolumn{1}{c}{Economic Modelling} \\
    \midrule
    \midrule
    Risk Aversion & +     & \multicolumn{1}{c}{+} & \multicolumn{1}{c}{+} \\
    Credit Risk & +     & \multicolumn{1}{c}{+} & \multicolumn{1}{c}{+} \\
    Operating Cost & +     & \multicolumn{1}{c}{+} & \multicolumn{1}{c}{+} \\
    Bank Size & NA    & \multicolumn{1}{c}{-} & \multicolumn{1}{c}{-} \\
    Liquidity Ratio & -     & N/A & N/A \\
    Management Quality & NA    & \multicolumn{1}{c}{-} & \multicolumn{1}{c}{-} \\
    Implicit Interest Payment & NA    & \multicolumn{1}{c}{+} & \multicolumn{1}{c}{+} \\
    Deposits Growth & NA    & N/A & N/A \\
    Operation Diversity & NA    & N/A & N/A \\
    Market Concentration & x     & \multicolumn{1}{c}{+} & \multicolumn{1}{c}{+} \\
    Real GDP Growth & +     & N/A & \multicolumn{1}{c}{-} \\
    Inflation & +     & N/A & \multicolumn{1}{c}{+} \\
    Ownership & Foreign (+) & N/A & N/A \\
    \midrule
    \midrule
    Sample & 80 Countries & OECD Countries & \makecell{EU Member \& \\  Candidates Countries } \\
    Estimation Methods & Pooled WLS & FE GLS & Pooled OLS \\
    \bottomrule
    \bottomrule
    \end{tabular}%
  \label{tab:addlabel}%
\end{table}%

% \newpage
% \clearpage
\singlespacing
\bibliography{\bib}
%\appendix
\end{document}